\documentstyle[preprint,prd,aps]{revtex}
\begin{document}
\draft
\preprint{
\vbox{
\halign{&##\hfil\cr
       AS-ITP-99-07\cr\cr}} }

\title{A Study of Gluon Propagator on Coarse Lattice}
\author{J.P. Ma}
\address{Institute of Theoretical Physics,\\
Academia Sinica, \\
P.O.Box 2735, Beijing 100080, China}

\maketitle 

\begin{abstract}
We study gluon propagator in Landau gauge with lattice QCD, where we use an
improved lattice action. The calculation of gluon propagator is performed on
lattices with the lattice spacing from $0.40$fm to 0.24fm and with the lattice
volume from $(2.40{\rm fm})^4$ to $(4.0{\rm fm})^4$. We try to fit our
results by two different ways, in the first one we interpret the calculated
gluon propagators as a function of the continuum momentum, while in the second we
interpret the propagators as a function of the lattice momentum. In the both we
use models which are the same in continuum limit. A qualitative 
 agreement between two fittings is found.
\end{abstract}
\vskip20pt

\preprint{\ \vbox{
\halign{&##\hfil\cr AS-ITP-99-00 \cr\cr}} }

\vspace{-5mm}

\vskip20pt

\vfill \eject
\narrowtext
\noindent 
{\bf 1. Introduction} \vskip 15pt Lattice QCD enables us to study the
nonperturbative nature of QCD from the first principle. As space-time is
replaced with a discreted lattice, physical results from Monte-Carlo
simulations can be obtained if the effect of the finite lattice spacing $a$
is under control. For this purpose simulations of lattice QCD are performed 
on large and fine lattices to ensure that the effect is small
enough. This is expensive. Using an improved action to
simulate lattice QCD on small and coarse lattices, it is possible 
to obtain physical results with less expense. The latter was proposed
long time ago\cite{SY}. It was suggested that one can use perturbative
theory to improve a lattice action and the effect of the finite lattice
spacing $a$ can systematically be removed. However, for lattice QCD
perturbative series converge so slowly that the improvement is not
significant. It was pointed out recently that perturbative theory can be made 
more effective by employing the tadpole improvement\cite{LM}. With an
improved action implemented with the tadepole improvement for lattice QCD 
some physical results are already obtained from simulations on a coarse
lattice. For example, hadron masses are well determined\cite{L1,L2,FW,Ma}.
\par
In this paper we study gluon propagator in Landau gauge with an improved
action on coarse lattices, where quarks are neglected. Perturbatively a
gluon is expected to behave like a massless particle. Because of the
color confinement of QCD such expectation can not be correct in a real world. A
nonperturbative study of gluon propagator is required and can help us to
understand the color confinement. Previous studies\cite{MMSM,CB,MM,DJA,DJA1} 
are undertaken with the
Wilson action, i.e., an unimproved action, they already show that gluon
propagator is more complicated than that from perturbative theory.
Especially, in Landau gauge it is well known that gluon propagator is
infrared finite due to Gribov's copies\cite{GR}, this is also 
proven to be true on lattice\cite{ZW}.
\par
We study gluon propagator on a series of lattices, of which the lattice
spacing $a$ is from $0.40$fm and to $0.22$fm. At $a=0.40$fm we also
investigate gluon propagator with lattice volume from $(2.40{\rm fm})^4$ to 
$(4.0{\rm fm})^4$. This allows us to study the effect of the finite lattice
spacing and of the finite volume. The whole calculation is performed with a
PC running at 400MHz.
\par
In continuum the gluon propagator $D_{\mu\nu}^{ab}$ in Landau gauge is
characterized by a single function $D(q^2)$: 
\begin{equation}
D_{\mu\nu}^{ab} (q) = (\delta _{\mu\nu}-\frac {q_\mu q_\nu}{q^2}
)\delta_{ab} D(q^2).
\end{equation}
The index $a$ and $b$ is the color index. We define a lattice version of the
function $D(q^2)$, which is denoted as $D_L$, and calculate it. We find that
the function $D_L$ can be well interpreted as a function of $q^2$ for 
$q^2$ up to3.5${\rm (GeV)^2}$.  We fit our results with a model proposed
in \cite{MMSM}. As our lattice spacing is rather larger, the effect of the 
finite lattice spacing needs to be studied in detail. 
For this we also try to fit our
propagator as a function of a modified lattice momentum variable, where the
same model is used with a slight modification suggested by perturbative
theory. An qualitative agreement is found between these two fittings.
\par
Our paper is organized as the following: In Sect. 2 we introduce the action
and our notation, we discuss different interpretations of
our data. In Sect.3 we fit our results by taking $D_L$ as a function
of $q^2$ with the model in \cite{MMSM}. In Sect.4 we fit our results
by taking $D_L$ as a function of lattice momentum with the same model, but
with a slight modification as suggested
by perturbative theory. In the both
sections we will discuss the effect of the finite lattice spacing and of the
finite volume on our fitting results. Sect.5 is our conclusion.
\par
\vskip 20pt 
\noindent 
{\bf 2. The Action and Our Notations}
\par
We take the one-loop improved action for gluon\cite{LW}, where the action
consists of plaquette, rectangle and paralellogram terms and is accurate up
to errors of $O(\alpha_s^2 a^2, a^4)$. Implementing tadpole improvement the
action becomes\cite{L1} 
\begin{eqnarray}
S(U) &=& \beta \sum_{pl} \frac{1}{3}{\rm Re Tr} (1-U_{pl}) +
\beta_{rt}\sum_{rt} \frac{1}{3} {\rm ReTr} (1-U_{tr}) +\beta_{pg} \sum_{pg} 
\frac{1}{3} {\rm ReTr} (1-U_{pg}), \cr \beta_{rt}&=& -\frac{\beta}{20 u_0^2}%
(1+0.4805 \alpha_s), \ \ \ \beta_{pg}=-\frac{\beta}{u_0^2}0.03325 \alpha_s, %
\cr u_0 &=& (\frac{1}{3}{\rm ReTr}\langle U_{pl}\rangle )^{\frac{1}{4}}, \ \
\ \alpha_s =-\frac {\ln(\frac{1}{3}{\rm ReTr}\langle U_{pl}\rangle )}{3.06839%
}
\end{eqnarray}
We used this action to generate gluonic configurations at $\beta=6.8$, $%
\beta=7.1$ and $\beta=7.4$, where the lattice size is $6^4$, $8^4$ and $10^4$
respectively. The parameter $u_0$ is determined by self-consistency. We also
generate configurations at $\beta=6.8$ with lattice size of $8^4$ and $10^4$%
. The configurations are generated with the pseudo heat bath method\cite{CM}%
, where the three $SU(3)$ subgroups were updated 3 times in each overall
update step. Configurations are separated by 40 sweeps to ensure that they
are statistically independent. The configurations used in this work are
summarized in Table 1.
\par
\vskip20pt \centerline{\bf Table 1} \vskip20pt

\begin{center}
\begin{tabular}{|r|r|r|r|r|}
\hline
\ \ \  $\beta$\ \  \ & \ \ \ \ \ $L^4$ \ \ \  & \ \ \ \ \ No. \ \  & \ $a$
\ \ \  & \ \ \ \ \ $u_0$ \ \ \ \  \\ \hline
\ \ 6.8\ \ \ \  & $6^4$\ \ \ \  & 500\ \ \ \  & 0.40fm\ \ \ \  & 0.8267\ \ \ \ \\ \hline
\ \ 6.8\ \ \ \  & $8^4$\ \ \ \  & 200\ \ \ \  & 0.40fm\ \ \ \  & 0.8267\ \ \ \ \\ \hline
\ \ 6.8\ \ \ \  & $10^4$\ \ \   & 200\ \ \ \  & 0.40fm\ \ \ \  & 0.8267\ \ \ \ \\ \hline 
\ \ 7.1\ \ \ \  & $8^4$ \ \ \   & 500\ \ \ \  & 0.33fm\ \ \ \  & 0.8434\ \ \ \ \\ \hline
\ \ 7.4\ \ \ \  & $10^4$\ \ \   & 250\ \ \ \  & 0.24fm\ \ \ \ & 0.8631\ \ \ \ \\ \hline
\end{tabular}
\end{center}
\par
\vskip15pt In Table 1 the lattice spacing $a$ is determined by the string
tension\cite{L1}. We use the naive steepest descent methode to fix the gauge
of our configurations \cite{CT}. We define: 
\begin{eqnarray}
\Delta(x) &=& \sum_\nu [ U_\nu (x-a\nu)-U_\nu(x)-h.c -{\rm trace\ term}], %
\cr \theta & =& \frac{1}{3V} \sum_x {\rm Tr} (\Delta (x) \Delta^\dagger (x)
),
\end{eqnarray}
where $V=L^4$ is the lattice volume. By gauge transformation we try to
minimize the quantity $\theta$. $\theta =0$ corresponds to the Landau gauge.
Numerically we require $\theta < 10^{-8}$. The gauge field is defined as 
\begin{equation}
A_\mu (x+a\hat\mu /2) = \frac {1}{2iag_0} (U_\mu (x) -U_\mu^\dagger (x)) - 
\frac{1}{6iag_0} {\rm Tr} (U_\mu (x) -U_\mu^\dagger (x)).
\end{equation}
It should be noted that in the Wilson action the bar coupling $g_0$ is
related to the $\beta$ as $g_0\beta =6$. With the action in Eq.(2) the bar
coupling $g_0$ is related to $\beta$ as 
\begin{equation}
g_0^2 = \frac {10}{\beta},
\end{equation}
if one neglects the tadpole improvement and the correction in the action at
one-loop level. Adding the improvement and the correction the relation
becomes complicated. In Landau gauge the quantity $\sum_{{\bf x}} A_0(t,{\bf %
x})$ is independent on $t$ for each configuration. 
We checked this under the numerical condition
for $\theta$. The quantity varies at different $t$ under $0.1\%$.
\par
The Fourier transformed gauge field is 
\begin{equation}
A_\mu ( q ) =a^4 \sum_x e^{-i q \cdot (x+a\hat\mu/2)} A_\mu (x),
\end{equation}
where $q$ is the momentum: 
\begin{equation}
q_\mu=\frac {2\pi}{L}n_\mu, \ \ \ \ n_\mu=0,1,\cdots, L-1.
\end{equation}
We define a function on lattice as 
\begin{equation}
D_L( q )= \frac {g_0^2 }{12V} \sum_\mu <0\vert {\rm Tr }( A_\mu ( q)
A_\mu^\dagger (q ) ) \vert 0> .
\end{equation}
\par
In the limit of $a\rightarrow 0$ the function $D_L(q)$ will approach to the
function $D(q^2)$ in the sence that it will be proportional to $D(q^2)$.
We will calculate the function $D_L(q)$ with the configurations listed
in Table 1. For further convenience we introduce 
\begin{equation}
\hat q_\mu =\frac{2}{a} \sin (\frac{1}{2} aq_\mu),\ \ \ \hat q^2=\sum_\mu 
\hat q_\mu^2,\ \ \ \ \hat q^4 =\sum_\mu \hat q_\mu^4.
\end{equation}
\par
In general the calculated function $D_L( q)$ will depend on $\hat q^2$, $%
\hat q^4$ and other possible invariants on lattice. if $aq_\mu$ is small
enough, $D_L( q)$ will approximately depend only on the variable 
\begin{equation}
q^2 =\sum_\mu q^2_\mu.
\end{equation}
Previous calculation on large lattice shows that even
for $a^2q^2 <1$ $D_L(q)$ still can not be interpreted as a 
function of $q^2$, the data points for $D_L(q)$ with momenta directed 
along one of the four axes behave differently than those with momenta 
directed off axis and the difference is significant. 
This clearly indicates that the rotation
invariance is violated due to finite lattice spacing and the effect is large.
With a improved action
like the one in Eq.(2) the effect of the 
violation will be reduced. Our data for $D_L( q)$
from the configurations with $\beta =7.4$ is shown in Fig.1a, where $D_L( q)$
is plotted as a function of $q^2$, and data with same $q^2$ are averaged.
From the figure we see that the data
points behave well as a function of $q^2$ in a range from $a^2q^2=0$ to $a^2q^2=5.13$.
For $a^2q^2>5.13$ the effect due to the violation of the rotation invariance 
becomes significant, indicated by a small jump in our data around$a^2q^2=6.3$. 
This is clearly
shown by the data from our smallst lattice in Fig.1b, where a small jump 
is roughly at $a^2q^2=10.0$ or at $q^2 =2.5{\rm GeV^2}$, for $a^2q^2>8.0$ 
the data points become unregular. 
The similar behave is also found for the lattice with the
size $L=8$. We plot our data from the lattice with $\beta =7.1$ in Fig.1c.
We will try to fit our data lying between $q^2=0$ and the $q^2$, where the data
points begin to become unregular, 
by taking $D_L$ as a function of $q^2$.  
\par
However it may be wondered that the effect of the finite lattice spacing is
small in the above consideration. To study this we also try to fit our
results by taking $D_L(q)$ as a function of $\hat q^2$. The replacement of $%
q^2$ by $\hat q^2$ can be thought as kinematic correction. In Fig.2a we
plot $D_L( q)$ from the lattice with $\beta =7.4$ as a function of $\hat q^2$, 
where data points with adjacent momenta lying within $\delta a^2\hat q^2<0.0001$ 
are averaged. 
From Fig.2a one can see that the data is not well scaled by $\hat q^2$ and
this indicates that $D_L$ may depends on on other lattice invariants and the
kinematic correction needs to be modified. Perturbative calculation with
the action in Eq.(2) also shows that $D_L( q)$ does not only depend on $\hat %
q^2$ but also on $\hat q^4$. At the tree-level without the tadpole
improvement one can obtain\cite{PW} 
\begin{equation}
D_L( q)=\frac{g_0^2}{3}\{-\frac{a^2}{12}+3[ \frac{1}{\hat q^2} -\frac{a^2}{18%
}\cdot \frac{\hat q^4}{(\hat q^2)^2}]\}+O(a^4)
\end{equation}
where there is a constant term. 
With this in mind the kinematic correction above needs to be further
modified. We introduce a new variable: 
\begin{equation}
\hat q^2_L=\hat q^2 +\frac{a^2}{12} \hat q^4,
\end{equation}
and for $a\rightarrow 0$: 
\begin{equation}
\hat q^2_L=q^2 +O(a^4).
\end{equation}
\par
In Fig. 2b we plot the gluon propagator as a function of $\hat q^2_L$,
where data points with adjacent momenta lying 
within $\delta a^2\hat q^2_L<0.0001$ 
are averaged.
From
the figure one can see that $D_L$ can be interpreted as a function of $\hat q%
^2_L$, although there is still a small fluctuation among the data points, but
it is much better than interpreting $D_L$ as a function of $\hat q^2$. We
will also fit our results by taking $D_L$ as a function of $\hat q_L^2$. 
\par
It should be pointed out that only on-shell quantities calculated with an
improved action are improved. For an operator in general one needs to
improve it to remove effect of finite $a$. In our case it is complicated to
do so. The reason is that the primary field on lattice is the gauge link $%
U_\mu(x)$, certain
effect of the finite $a$ at order of $a^2$ is introduced by extracting
the gauge field $A_\mu (x)$ from $U_\mu(x)$, and some effect at
order of $a^2$ arises if one calculate the propagator. To obtain an improved 
$D_L( q)$ one should improve the extraction and then improve the operator $%
A_\mu ( q) A_\mu^\dagger ( q)$. In this work we will not consider to obtain 
an improved $D_L(q)$. 
\par
\noindent
\vskip20pt
\noindent  
{\bf 3. Fitting propagators as a function of continuum momentum}
\par
In this section we fit our calculated propagators as functions of $q^2$ with
the model\cite{MMSM} 
\begin{equation}
D_L(q^2) =\frac {Z}{(M^2)^{1+\alpha } + (q^2)^{1+\alpha } }
\end{equation}
where $M$ is a dimensional parameter and $\alpha$ is the anomalous dimension. 
$Z$ is a parameter related to $g_0$ and the renormalization constant of wave-function. 
In this and the next section we will not discuss this parameter. 
In the fitting of our propagators we exclude the data points with the lowest$
q^2$ including the point with $q^2=0$. We try to include more data points
with larger $q^2$ untile $\chi^2$ per degree of freedom close to 1, but not
larger than 1. 
Beside this the total fitting range is limited 
because of the reason discussed in the last section. Our fitting results
are summarized in Table 2: 
\par
\vskip20pt \centerline{\bf Table 2} \vskip20pt
\par
\begin{center}
\tabcolsep4pt
\begin{tabular}{|r|r|r|r|r|r|r|}
\hline
 $\beta$ &\ $L $ &\ Range \  & \  $Z$ \   & \
 $a^2M^2$ \  & \ $\alpha $ \   & \ $\chi ^2/dof$ \ \\ \hline 
\ 6.8\ &\ 8\   &\ 3-9\   &\ 5.811(133)\ 
&\ 1.654(92)\ &\ 0.107(24)\ & \ 0.35\ \\ \hline
\ 6.8\ &\ 10\  &\ 3-12\ &\ 5.975(77)\  
&\ 1.730(49)\ &\ 0.111(14)\ &\ 0.07\   \\ \hline
\ 7.1\ &\ 8\ &\ 3-12\ &\ 5.511(14)\ 
&\ 0.949(106)\ &\ 0.187(18)\ &\ 0.28\ \\ \hline
\ 7.4\ &\ 10\ &\ 3-14\ &\ 5.298(75)\  
&\ 0.602(41)\ &\ 0.312(13)\ &\ 0.08\   \\ \hline
\end{tabular}
\end{center}
\par
\vskip20pt
The data points are labeled by an integer $n$, smaller $n$  corresponds 
to smaller $q^2$, $n=1$ corresponds to $q^2=0$. 
The range is the fitting range and given by this integer. 
We do not list our results for the configurations with $
\beta=6.8,L=6$, because the fitting quality is poor with small number of
data points. We also tried to fit our data with other models. For example, 
the model proposed in \cite{DJA}, but it can not well describe our data
in the range given above by the indication of large $\chi^2/dof$. 
To illustrate our fitting quality we plot our data with fitting results 
in Fig.3a and in Fig.3b for $\beta =7.4$ and for $\beta =6.8$ respectively. 
\par
From our fitting results and with the lattice spacing given in Table 1 
we obtain the dimensional parameter $M$ in GeV:
\begin{eqnarray} 
M &=& 0.633(28)\ \ {\rm for\ \beta =6.8, L=8},\ \ \ M=0.647(14)\ \  
                   {\rm for\ \beta =6.8, L=10}, \cr
M &=& 0.582(56)\ \ {\rm for\ \beta =7.1, L=8},\ \
M=0.636(34)\ \ {\rm for\ \beta =7.4, L=10}. 
\end{eqnarray}
We note that the values of $M$ at $\beta =6.8$ with $L=8$ and with 
$L=10$ can be taken as the same within the errors, this may 
indicate that the effect of the finite volume is small. However, 
we have only two values from two lattices, where the difference in volume
is not large, a study on larger lattices with different volumes 
is needed to confirm the above conclusion. From above the values of $M$ 
remain relatively stable with different lattice spacings $a$.  
We use a standard weighted least-squares procedure to average the $M$-
values and obtain the central value $M_c$:
\begin{equation} 
M_c=641(11){\rm MeV}.
\end{equation} 
We also note that the deviation of $M$ at $\beta=7.1$ from $M_c$ is
the largest, but it is still in the range of the errors. The deviation
of other values is small. This leads to conclude that there is no 
significant dependence of $M$ on the finite $a$ and 
to interpret the central value $M_c$ as the parameter $M$ 
in the continuum limit.  
\par 
As the next we study the anomalous dimension $\alpha$. From the fitting results 
in Table. 2, the values from two lattices with different volumes at $\beta =6.8$ 
are the same within the errors. This may also indicate that the effect 
of the finite volume in $\alpha$ is small as that in $M$. 
For values at different $\beta$ there is 
clearly a $a$-dependence of $\alpha$. We try to model the dependence with a
simple formula:
\begin{equation} 
\alpha (a) =\alpha_0 +\alpha_1 a^2 
\end{equation}
where $a$ is in unit of fm and $\alpha_0$ can be thought as the anomalous dimension 
in continuum. At $\beta =6.8$ we take the value of $\alpha$ with $L=10$ for the fit.
The results is
\begin{equation}
\alpha_0=0.421(20),\ \ \alpha_1=-1.97(19),\ \ \ \chi^2=1.42. 
\end{equation} 
In Fig. 4 we illustrate the quality of this fit. The value of $\chi^2$ is acceptable. 
With the above discussion we conclude in this section with our data that 
the gluon propagator in continuum has a mass scale at $641(11)$MeV and there is 
an anomalous dimension which is 0.421(20). 
\par
\vskip20pt
\noindent
{\bf 4. Fitting propagator as a function of lattice momentum}
\par
In this section we fit our data with the same model as in the last section, 
where the variable $q^2$ is replaced by $\hat q_L^2$, and a constant term is added
as suggested by perturbative theory. This term will vanish in the limit
of $a\rightarrow 0$. The model with such modification is: 
\begin{equation}
\label{ML1} 
D_L(q^2) =\frac {Z}{(M^2)^{1+\alpha } + (\hat q_L^2)^{1+\alpha } }+ca^2.
\end{equation} 
\par
The data selection for fitting is similar as in the last section. We discarded
the first two points and try to include more data points in the direction 
of increasing $\hat q_L^2$, until $\chi^2$ per degree of freedom close 
to 1 but never larger than 1. Our fitting results are summarized
in Table 3. 
\vskip20pt \centerline{\bf Table 3} \vskip20pt
\par
\begin{center}
\tabcolsep4pt
\begin{tabular}{|r|r|r|r|r|r|r|r|}
\hline
\ $\beta$\ &\ $L$ & \ Range \  & \  $Z$ \   & \
 $a^2M^2$ \ & \ $\alpha $ \ & \ $c$\ \ \ &\ $\chi ^2/dof$ \ \\ \hline
\ 6.8\ &\ 6\   & 3-12\ & 6.297(72)\ 
&\ 1.433(67)\  &\ -0.019(8)\ & -0.319(11)\ & 0.65 \ \\ \hline 
\ 6.8\ &\ 8\  & 3-13\  & 6.204(68)\ 
&\ 1.496(56)\  & -0.080(10)\ & -0.379(14)\ & 0.50 \ \\ \hline
\ 6.8\ &\ 10\ & 3-16\  & 6.007(59)\ 
&\ 1.440(46)\ & -0.097(11)\ & -0.365(14)\ & 0.44 \  \\ \hline
\ 7.1\ &\ 8\ & 3-13\  & 6.203(69)\ 
&\ 0.945(39)\ & 0.026(12)\ & -0.383(15)\ & 0.76 \ \\   \hline
\ 7.4\ &\ 10\ & 3-15\ & 5.809(56)\ 
&\  0.546(20)\ & 0.147(13)\ & -0.361(16)\ & 0.58 \ \\ \hline
\end{tabular}
\end{center}
\par
\vskip20pt
The data points are labeled by an integer $n$, smaller $n$  corresponds 
to smaller $q_L^2$, $n=1$ corresponds to $\hat q^2_L=0$. 
The range is the fitting range and is given by this integer.
To illustrate out fitting we plot our data with the fitted curves 
in Fig.5a and Fig.5b as examples.  
From above one can realize that the $c$ roughly remain as the same at different 
$\beta$ and at different lattice volumes except the one at
$\beta =6.8,L=6$. For other values of $c$ the variation is 
in the range within errors. This fact 
will ensure that the term with $c$ in Eq.(19) will vanish in the limit 
of $a\rightarrow 0$. 
\par 
With the lattice spacing in Table 1 we obtain the mass parameter: 
\begin{eqnarray}
M &=& 0.590(11)\ \ {\rm for\ \beta =6.8, L=6},\cr 
M &=& 0.602(9)\ \ {\rm for\ \beta =6.8, L=8},\ \ \ M=0.591(8)\ \  
                   {\rm for\ \beta =6.8, L=10}, \cr
M &=& 0.580(12)\ \ {\rm for\ \beta =7.1, L=8},\ \
M=0.606(42)\ \ {\rm for\ \beta =7.4, L=10}. 
\end{eqnarray}
We observe that the value of $M$ is within the errors stable in the range of $\beta$ and in the 
range of the volume of our lattices. This also indicates as in the last section 
that the gluon propagator in continuum limit contains a mass scale. We calculate with the 
data the central value: 
\begin{equation} 
M_c=592(5){\rm MeV}. 
\end{equation} 
This value is $10\%$ smaller than the one in the last section. The reason 
may be due to the difference between the variables $q^2$ and $q^2_L$, i.e.,
the effect of the finite lattice spacing is still observable, it may be at 
order of $10\%$.   
\par 
At $\beta =6.8$ the parameter $\alpha$ from the lattice with the volume $L=6$ is quite different 
than the values from other larger lattices, while there is no large deviation 
between the values from $L=8$ and from $L=10$. The similar is also found 
with the parameter $c$. This may be explained that 
at $\beta =6.8$ the effect of the finite volume is 
significant for $L$ smaller than 8. We try to fit this dependence with the same
relation as in the last section:  
\begin{equation} 
\alpha  =\alpha_0 +\alpha_1 a^2,  
\end{equation}
the fitting results are:
\begin{equation}
\alpha_0 =0.285(20),\ \ \alpha_1 =-2.38(17),\ \ \chi^2=0.07. 
\end{equation} 
The fitting quality is better than that in the last section. But $\alpha_0$ 
is different than this in the last section. Here $\alpha_0$ may not
be interpreted as the anomalous dimension. The reason is: 
In the model in Eq.(14) the effect of the finite $a$ is only contained 
in the parameters, while in the model in Eq.(19) not only the parameters
implicitly contain possible effect of the finite $a$, but also some 
effect of the finite $a$ is explicitly included, representing 
by the term with $c$ and with the variable $\hat q_L^2$. If the latter 
can be neglected, one may interpret $\alpha_0$ 
as the anomalous dimension, which is $30\%$ smaller than $\alpha_0$ in 
the last section. This may be thought as a qualitative agreement.
\par
\vskip 20pt 
\noindent
{\bf 5. Conclusion}
\par 
In this work we studied gluon propagator with an improved action 
on lattice. Unlike with the Wilson action, we find that the effect 
due to the violation of the rotation invariance is reduced, resulting
in that $D_L$ can be well interpreted as a function of $q^2$ as 
a continuum variable for $a^2q^2$ not larger than 5.13. 
We fitted the gluon propagator with a model proposed in \cite{MMSM}
and studied the effect of the finite lattice spacing and of
the finite volume. Our conclusion is that in the continuum limit
the gluon propagator has a mass scale around 600MeV and a 
anomalous dimension which is 0.412. We also tried to study 
the gluon propagators as a function of an improved lattice 
momentum $q_L^2$ and fitted them with the model which is the same 
in the continuum limit as the one used before. The fitting results are 
qualitatively in agreement with the study of $D_L$ as a function 
of $q^2$. The difference between two fittings may be explained by the 
effect of the finite lattice spacing, and by the effect depending 
on how to model the gluon propagator $D_L$. Simulations with large lattices 
may be required  to study the difference in detail. 
\par

Finally it should be pointed that calculations with larger lattice than 
those in our work will allow us not only to study the effect of the finite
lattice spacing and of the finite volume in more detail but also 
to have more data points with small $q^2$.

\vskip15pt

\vfil\eject

\vfil\eject 
{\center{\bf Figure Caption}}
\par\noindent
Fig.1a: The gluon propagator $D_L$ at $\beta =7.4$ with $L=10$ 
plotted as a function of $q^2$. Statistical errors are not plotted,
they are small and around $1\%$. 
\par\noindent
Fig.1b: The gluon propagator $D_L$ at $\beta =6.8$ with $L=6$ 
plotted as a function of $q^2$.
\par\noindent
Fig.1c: The gluon propagator $D_L$ at $\beta =7.1$ with $L=8$ 
plotted as a function of $q^2$.
\par\noindent
Fig.2a: The gluon propagator $D_L$ at $\beta =7.4$ with $L=8$ 
plotted as a function of $\hat q^2$. The x-axis is for $a^2\hat q^2$.
\par\noindent
Fig.2b: The gluon propagator $D_L$ at $\beta =7.4$ with $L=8$ 
plotted as a function of $\hat q^2_L$. The x-axis is for $a^2\hat q^2_L$. 
\par\noindent 
Fig.3a: The gluon propagator $D_L$ at $\beta =7.4$ with $L=10$ 
plotted as a function of $q^2$, the line is from our fit. All data 
shown in this figure are used in fitting. 
\par\noindent
Fig.3b: The gluon propagator $D_L$ at $\beta =6.8$ with $L=8$ 
plotted as a function of $q^2$, the line is from our fit. All data 
shown in this figure are used in fitting.
\par\noindent
Fig.4 The extrapolation of $\alpha$. The x-axis is for $a^2$ in ${\rm (fm)}^2$, 
the y-axis is for $\alpha$. 
\par\noindent 
Fig.5a: The gluon propagator $D_L$ at $\beta =7.4$ with $L=10$ 
plotted as a function of $\hat q^2_L$, the line is from our fit with 
the model in Eq.(19). All data 
shown in this figure are used in fitting. 
\par\noindent 
Fig.5b: The gluon propagator $D_L$ at $\beta =6.8$ with $L=6$ 
plotted as a function of $\hat q^2_L$, the line is from our fit with 
the model in Eq.(19). All data 
shown in this figure are used in fitting. 
\par\noindent
Fig.6 The extrapolation of $\alpha$. The x-axis is for $a^2$ in (fm)$^2$, 
the y-axis is for $\alpha$. The $\alpha$ is defined in Eq.(19).
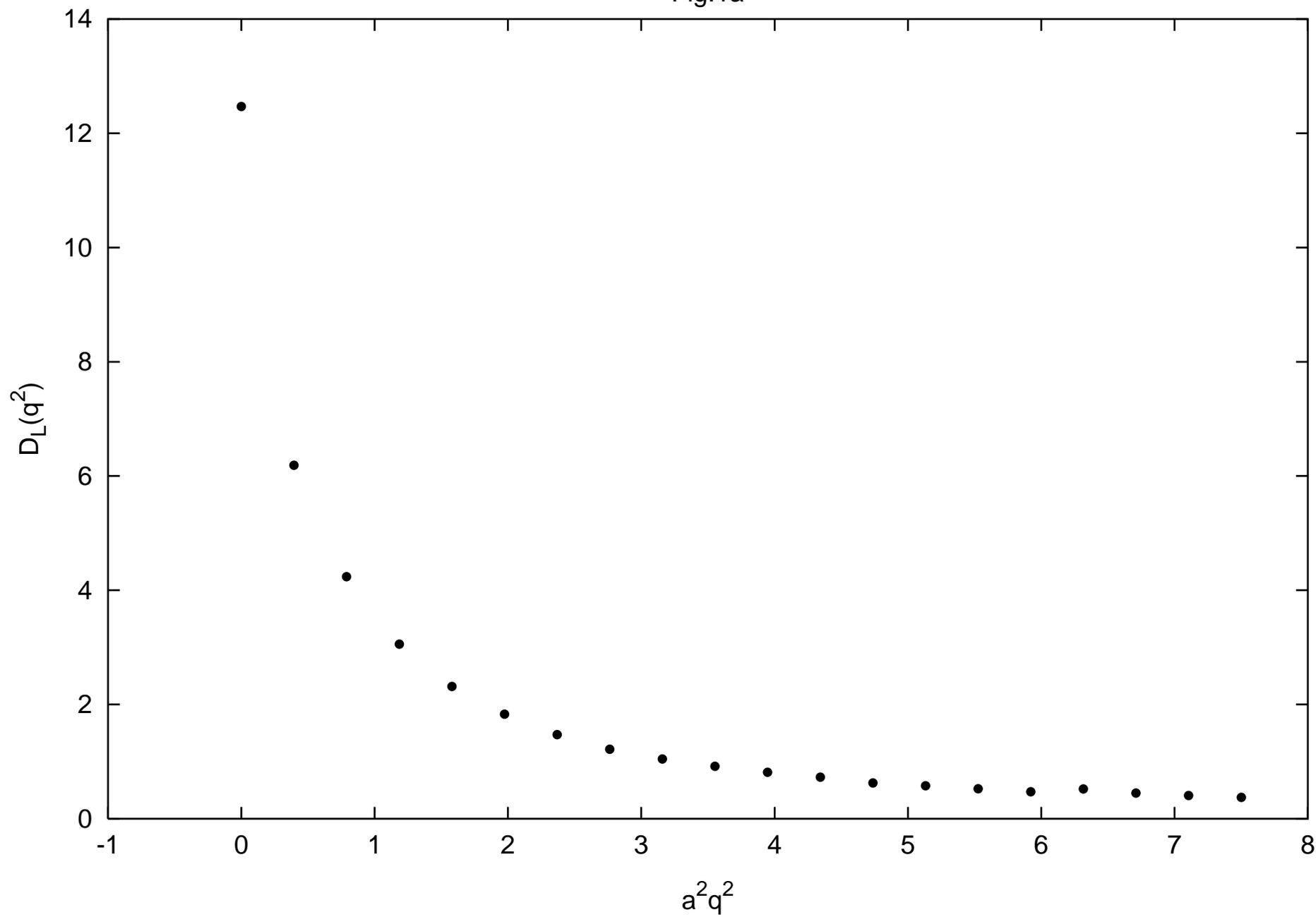
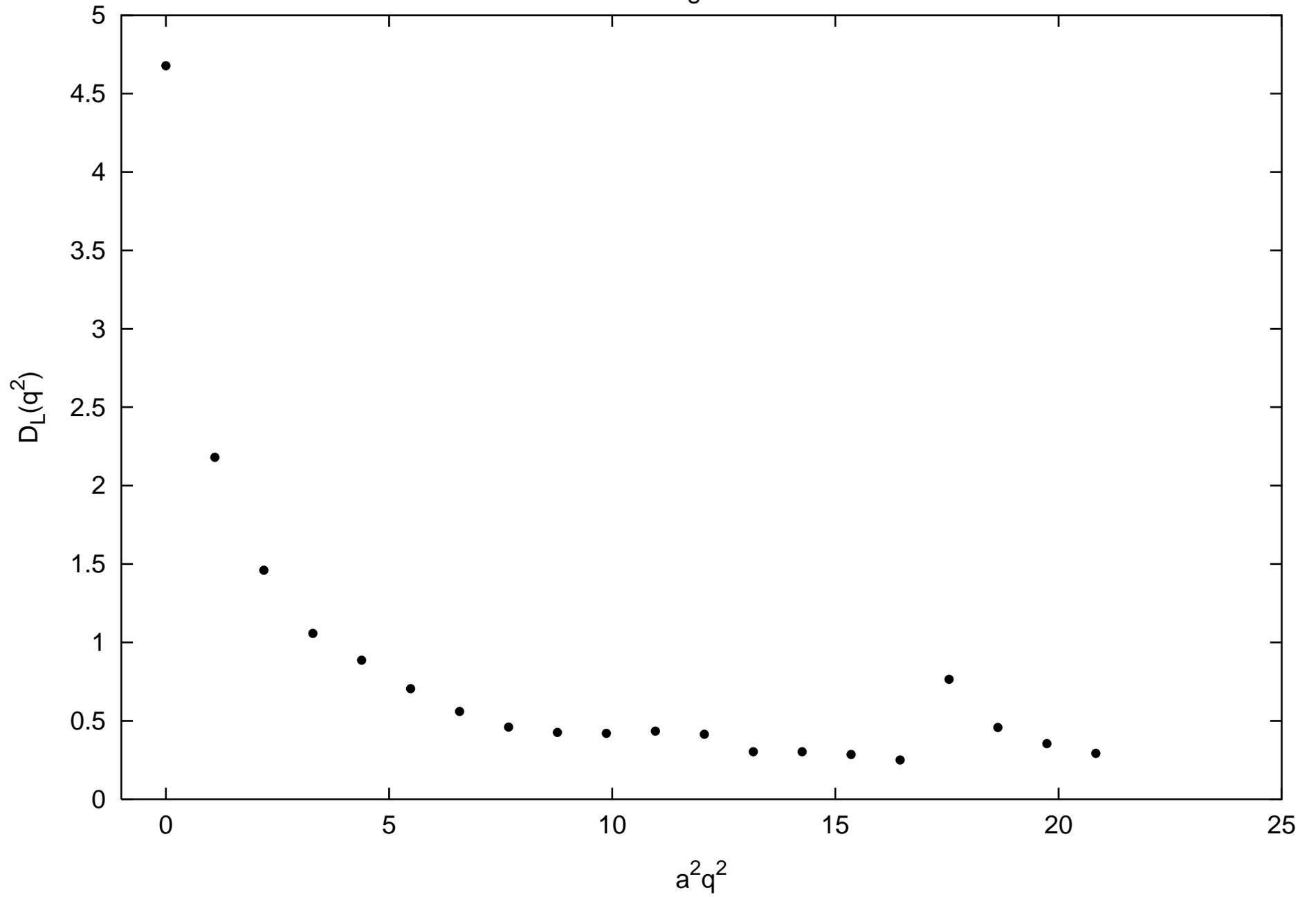
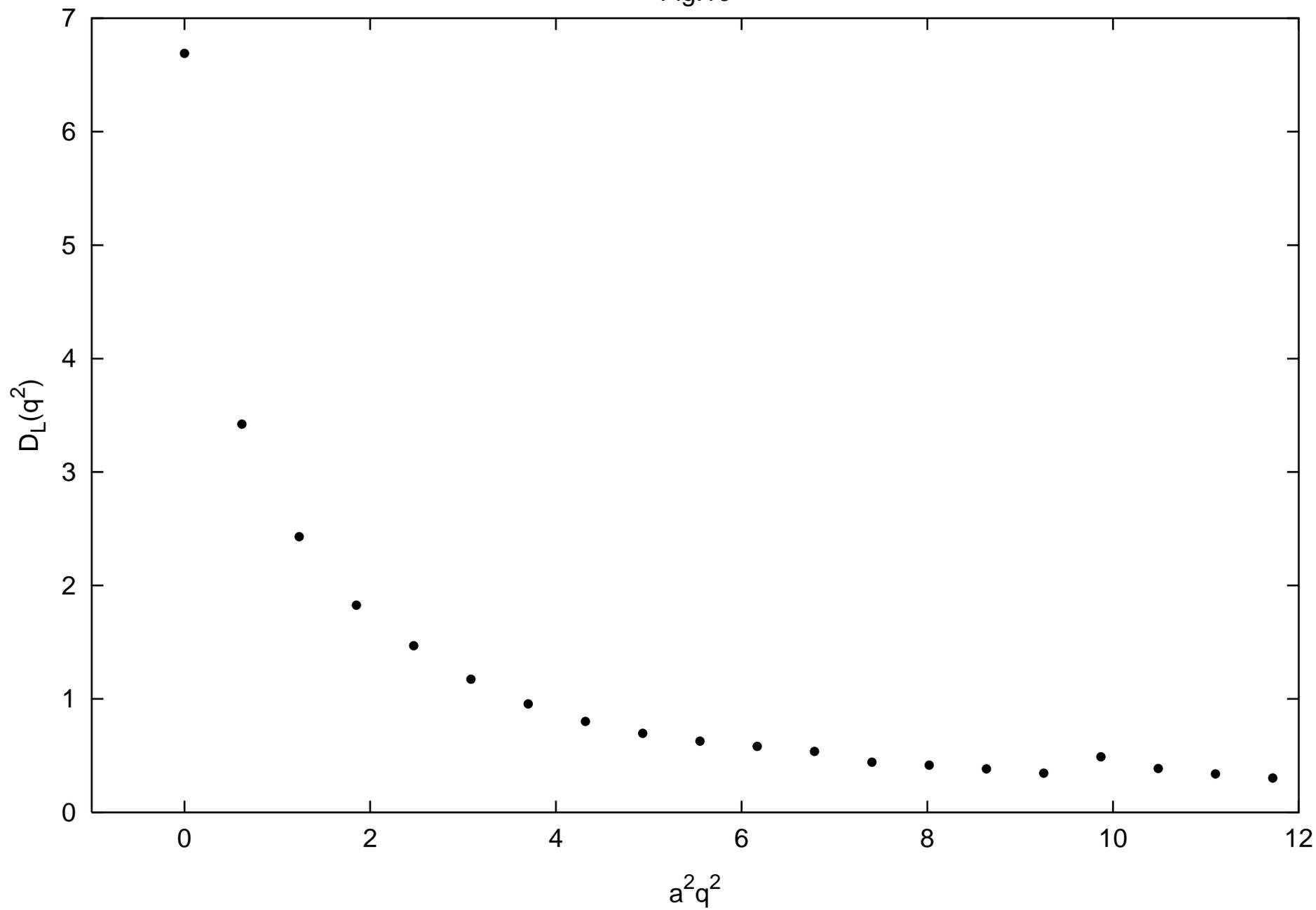
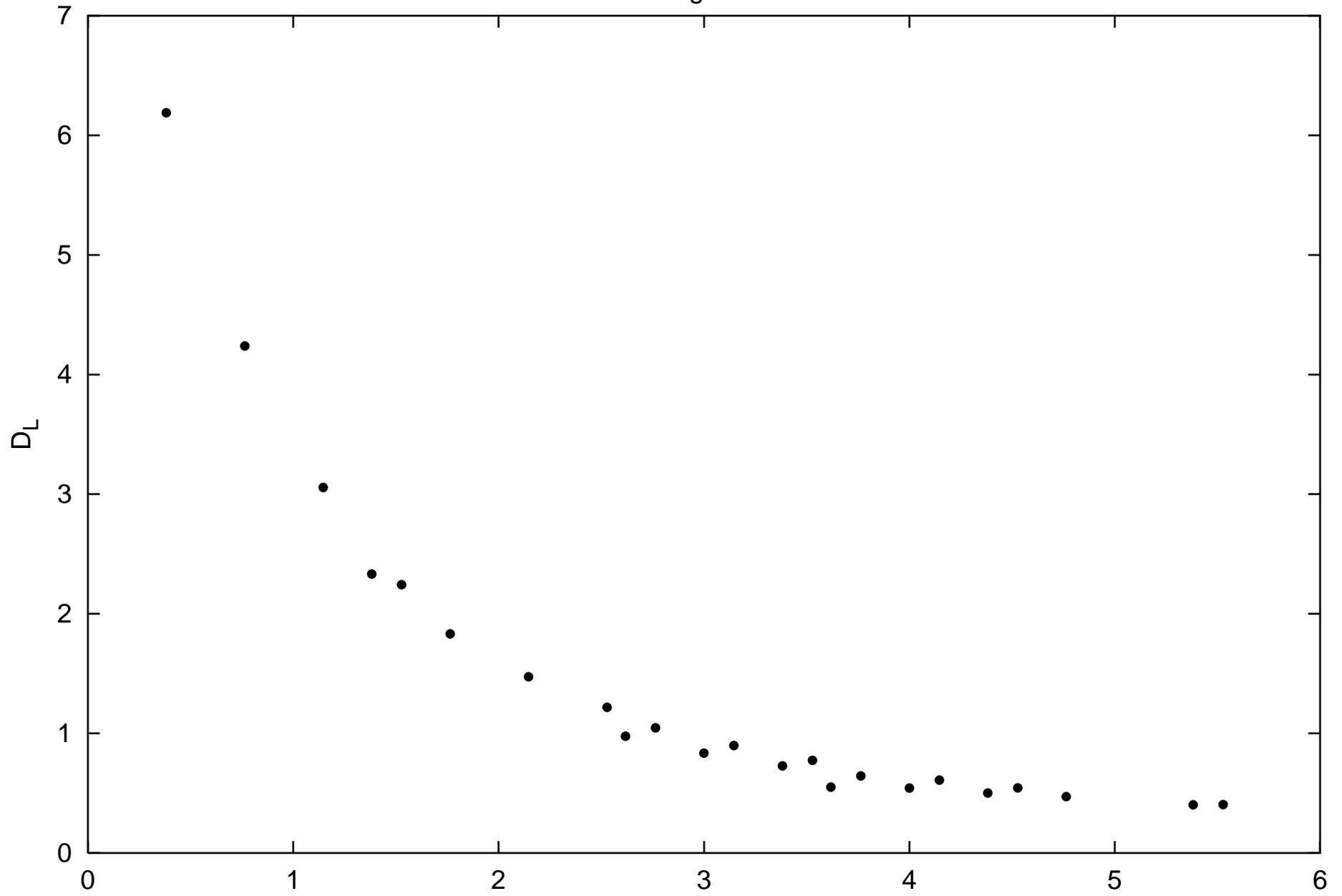
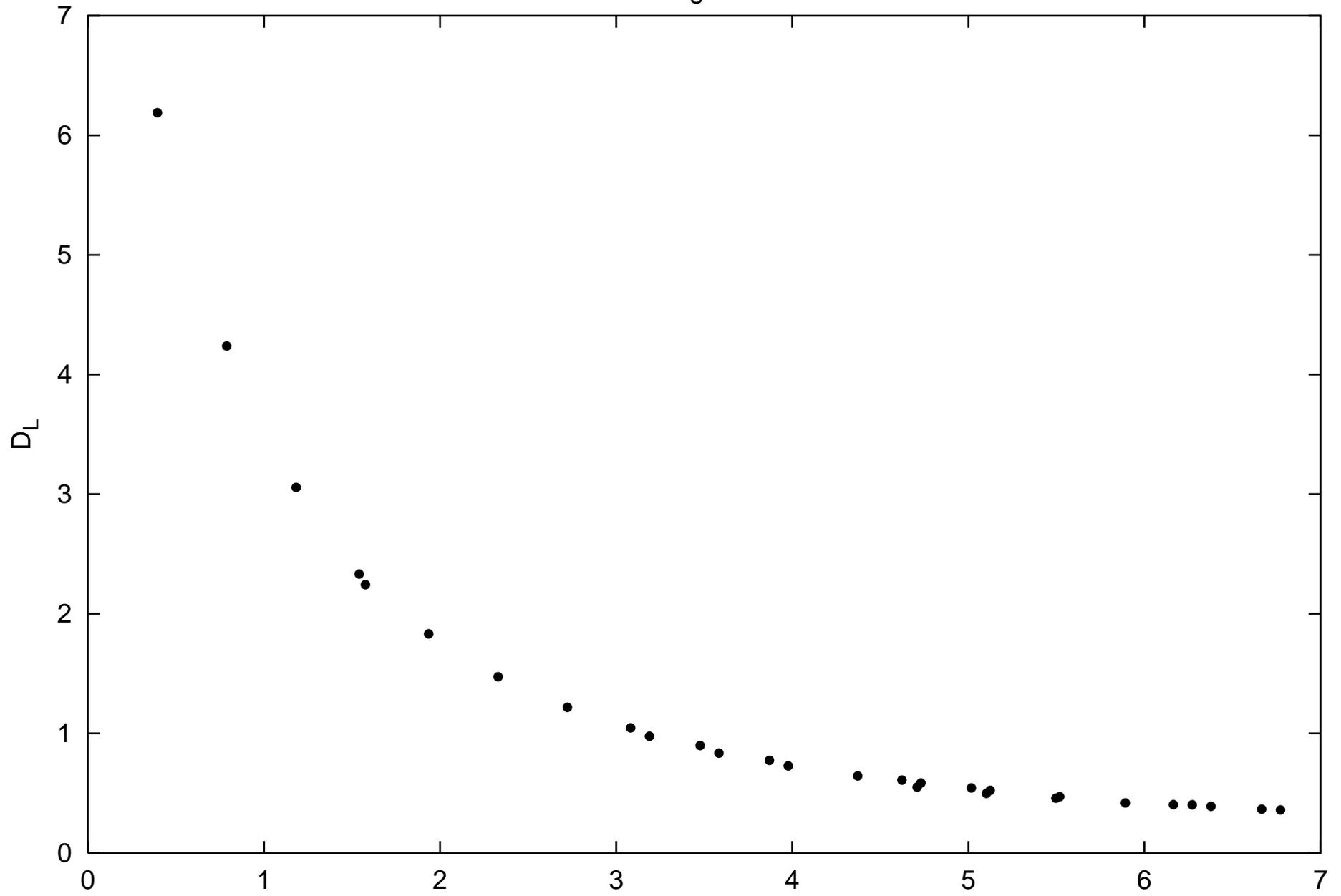
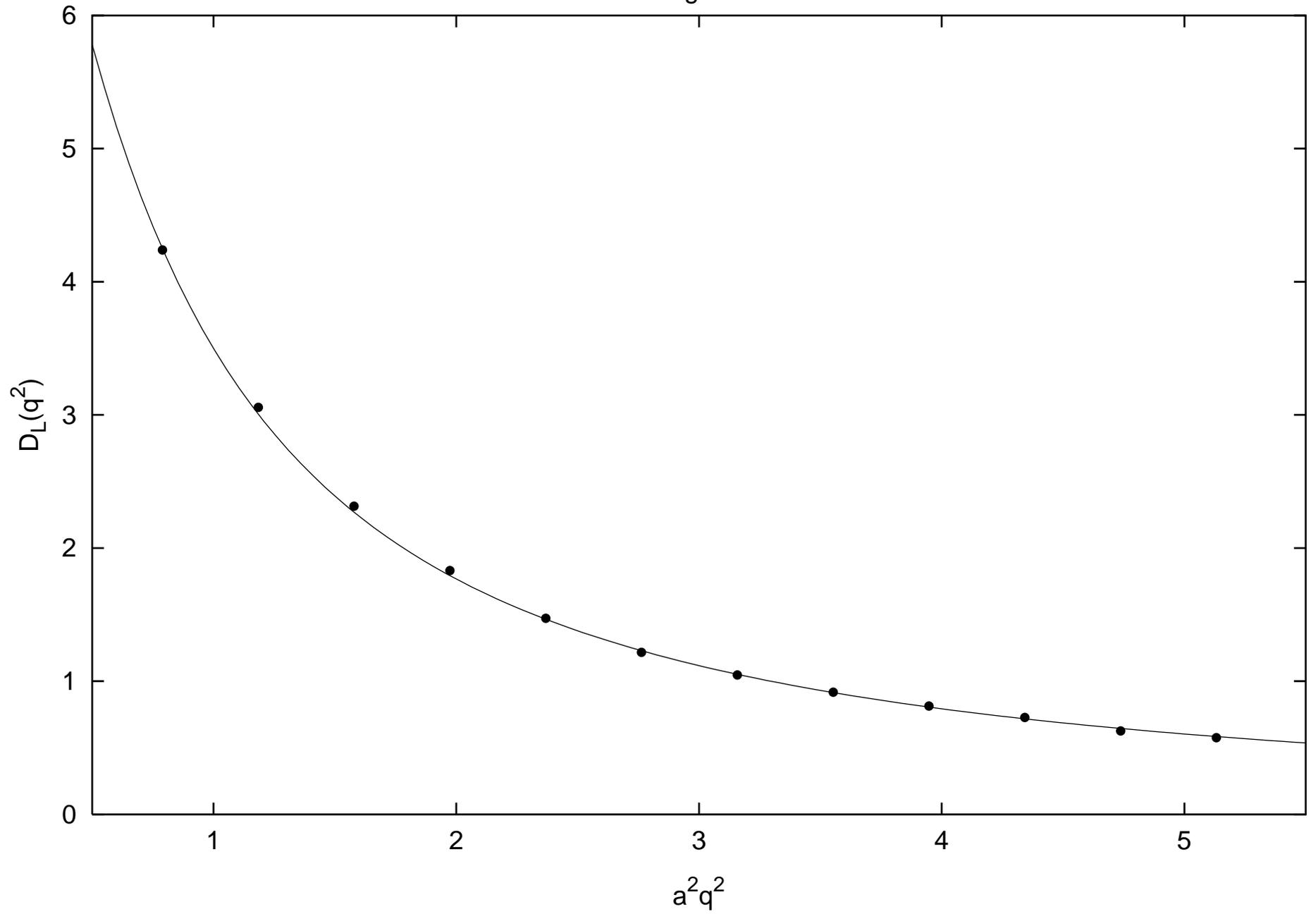
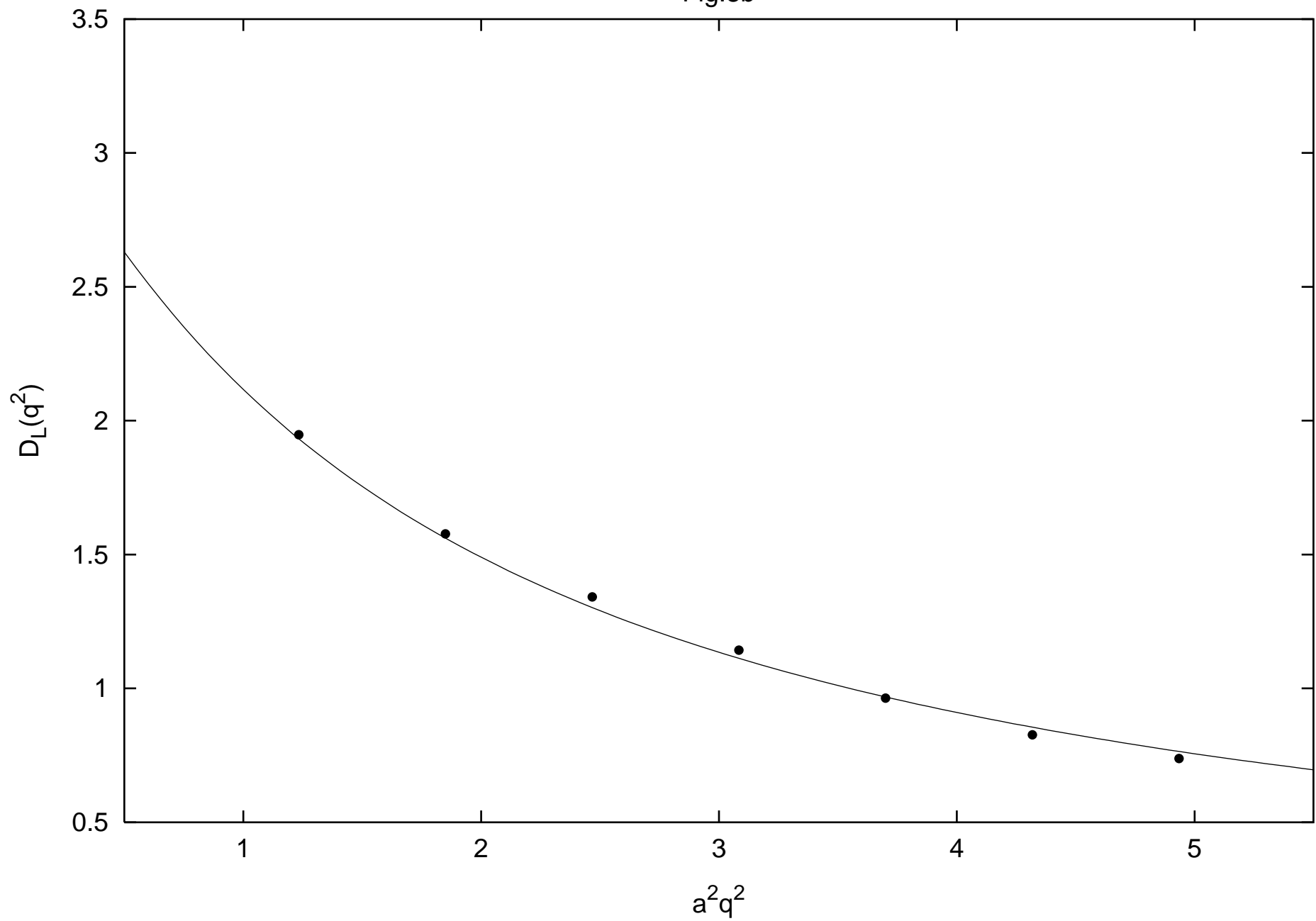
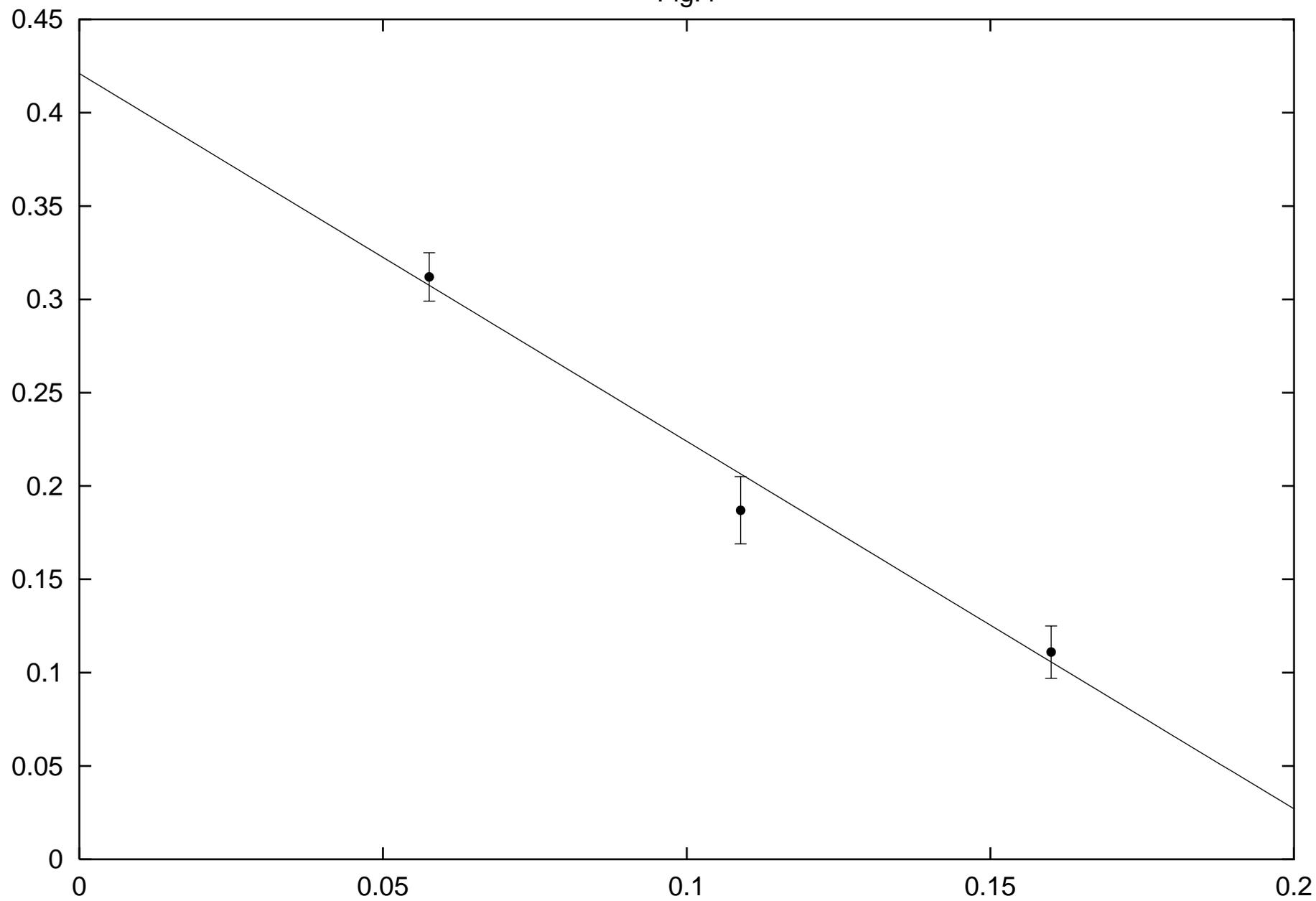
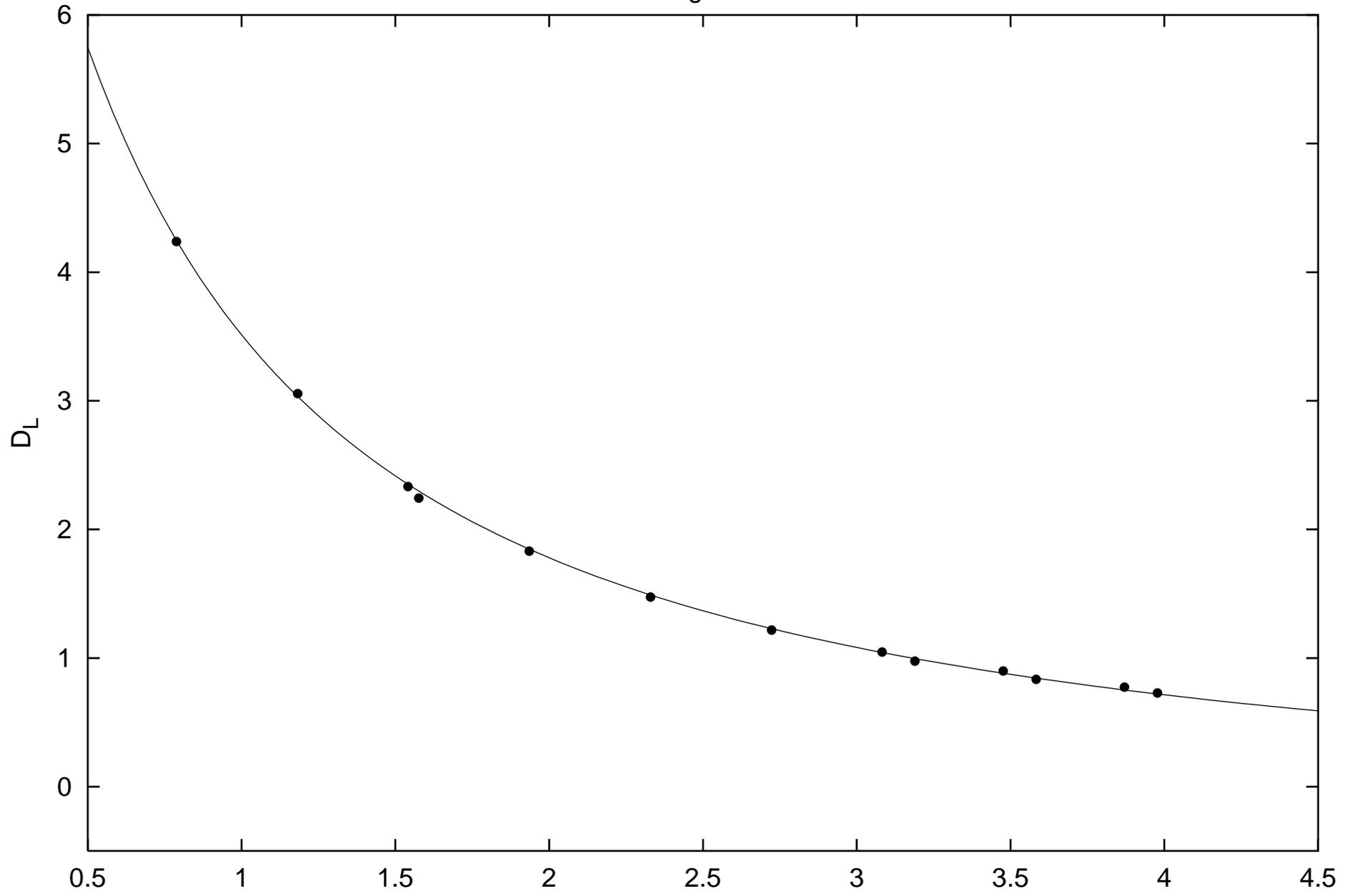
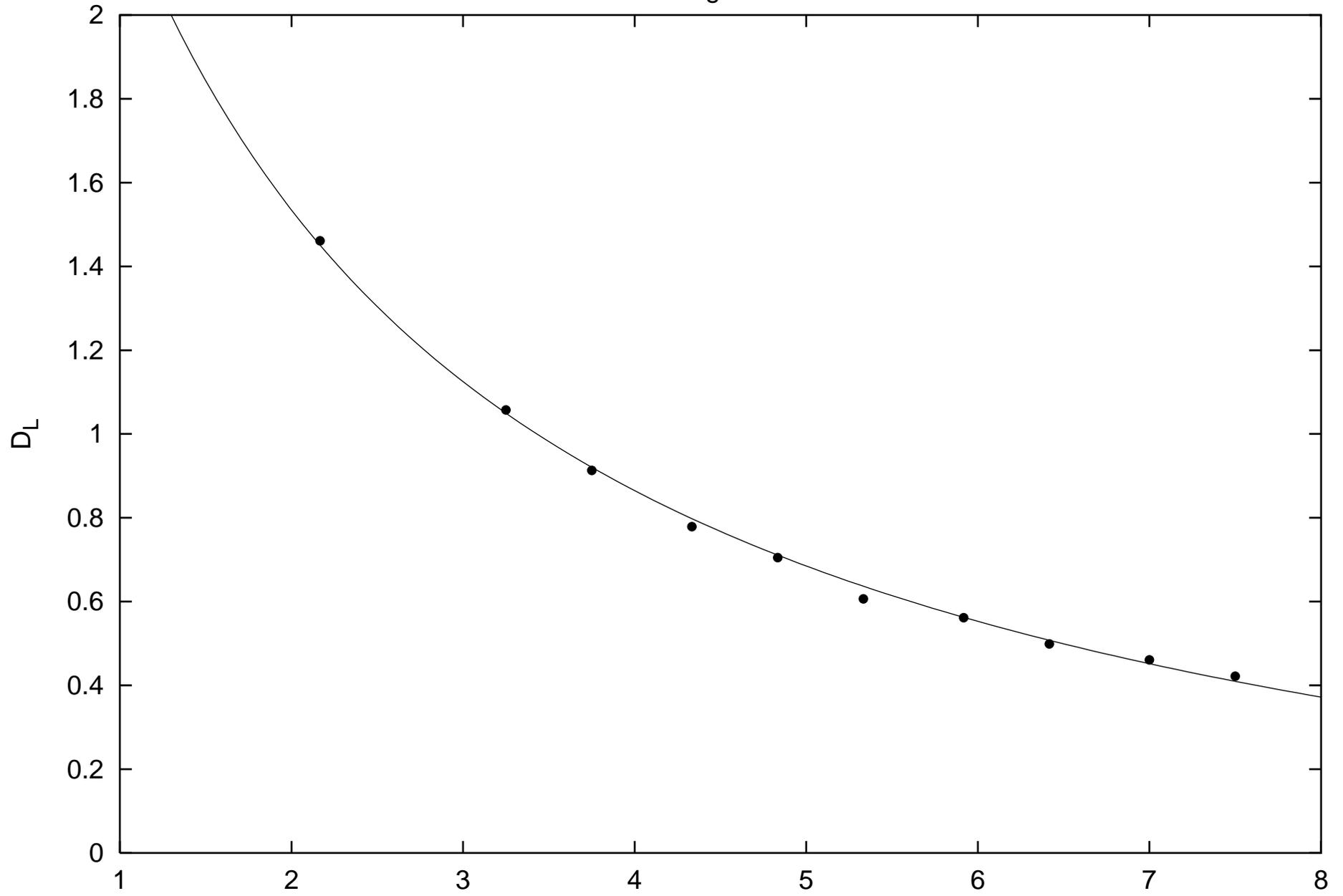
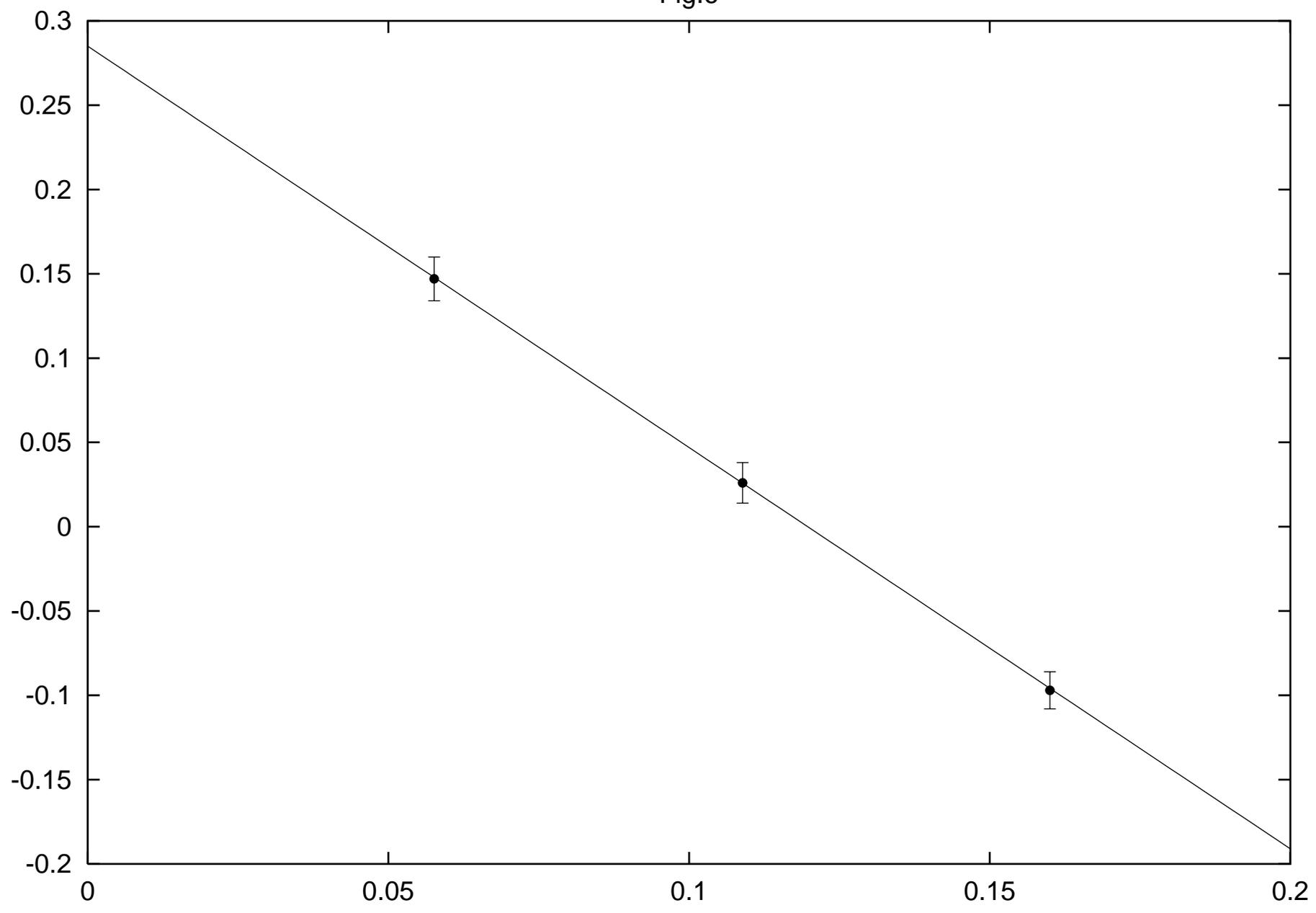
\end{document}